\documentclass[apj]{emulateapj}

\usepackage{amsmath} 
\usepackage{bm} 

\submitted{Accepted for publication in ApJ}

\shorttitle{On the tidal radius}
\shortauthors{Gajda \& {\L}okas}

\begin{document}

\title{On the tidal radius of satellites on prograde and retrograde orbits}
\author{Grzegorz Gajda and Ewa L. {\L}okas}
\affil{Nicolaus Copernicus Astronomical Center, Bartycka 18, 00-716 Warsaw, Poland}

\begin{abstract}
A tidal radius is a distance from a satellite orbiting in a host potential beyond which its material is stripped by
the tidal force. We derive a revised expression for the tidal radius of a rotating satellite which properly takes
into account the possibility of prograde and retrograde orbits of stars. Besides the eccentricity of the satellite orbit,
the tidal radius depends also on the ratio of the satellite internal angular velocity to the orbital angular velocity.
We compare our formula to the results of two $N$-body simulations of dwarf galaxies orbiting a Milky Way-like host on a
prograde and retrograde orbit. The tidal radius for the retrograde case is larger than for the prograde. We introduce a
kinematic radius separating stars still orbiting the dwarf galaxy from those already stripped
and following the potential of the host galaxy. We find that the tidal radius matches very well the kinematic radius. Our results provide a connection between the formalism of the tidal radius derivation and the theory of resonant stripping.
\end{abstract}

\keywords{galaxies: dwarf --- galaxies: interactions --- galaxies: kinematics and dynamics --- galaxies: structure}

\section{Introduction}

The tidal radius $r_\mathrm{t}$ is a theoretical boundary of a satellite object (e.g. a star cluster, a galaxy) beyond
which its material is stripped by the tidal forces originating from the potential of a larger host object (e.g.
another galaxy, a galaxy cluster). It was proposed for the first time by \citet{vhoerner57} in the context of globular
clusters orbiting around the Milky Way. A strict theoretical definition exists only for satellites on circular orbits,
for which the tidal radius is identical to the position of L1/L2 Lagrange points \citep[see][]{bt08}. \citet{king62}
argued that for an eccentric orbit of a satellite, a star located at the instantaneous tidal radius is
stationary in a rotating reference frame and is attracted neither towards the satellite nor towards the host. He also
argued that the satellites are truncated during pericenter passages to the size indicated by the pericentric tidal
radius.

However, already \citet{henon70} and \citet{keenan_innanen75} noticed that retrograde orbits in the restricted
three-body problem are stable up to much larger distances than prograde ones, exceeding the \citet{king62} tidal
radius. \citet{read06} derived an expression for $r_\mathrm{t}$ that takes into account three basic types of orbits:
prograde, radial and retrograde. Indeed, the tidal radius for retrograde orbits turned out to be larger than for
radial ones, which in turn was larger than for prograde orbits.

Interactions of galaxies with the potential of larger objects have proved capable of transforming their
morphology and kinematics on various scales: those of dwarf galaxies orbiting the Milky Way-like galaxy \citep{mayer01,
kazantzidis11}, galaxies in groups \citep{villalobos12} and clusters \citep{mastropietro05, bialas15}. One of the
factors influencing the final outcome of such a process is the initial inclination of the galaxy disk with respect to
its orbit \citep{villalobos12, lokas15}. The result of an encounter of equal-mass disk galaxies also depends on the
direction of their rotation \citep{holmberg41, toomre_toomre72}.

The tidal radius is important for studying globular clusters \citep{webb13} and dwarf galaxies \citep{lokas13}
as it can help disentangle
their bound parts from the material which has been already lost and forms tidal tails \citep{kupper10, kupper12}.
Semi-analytical models of galaxy formation also rely on comparisons of the tidal radius to the size of the galaxy
to describe timescales of stripping \citep{henriques_thomas10, chang13}.

In this paper we derive a new expression for the tidal radius, which properly takes into
account both the eccentric orbit of the satellite and the orientation of its internal rotation. We apply this
formula to $N$-body simulations of a disky dwarf galaxy orbiting a Milky Way-like host and compare its predictions to
features detected in mass and velocity distributions of the dwarf.

\section{Derivation}
\label{sec_derivation}

Let a dwarf galaxy with a spherically symmetric mass distribution $m(x)$ orbit around a host galaxy with a spherically
symmetric mass distribution $M(x)$. In a reference frame centered on the dwarf the
acceleration of a star belonging to the dwarf, located at $\bm{x}_{\mathrm{s}}$, is given by
\begin{multline}
\ddot{\bm{x}}_{\mathrm{s}}=-Gm(|\bm{x}_{\mathrm{s}}|)\frac{\bm{x}_{\mathrm{s}}}
{|\bm{x}_{\mathrm{s}}|^3}-GM(|\bm{x}_{\mathrm{s}}-
\bm{x}_{\mathrm{h}}|)\frac{\bm{x}_{\mathrm{s}}-\bm{x}_{\mathrm{h}}}
{|\bm{x}_{\mathrm{s}}-\bm{x}_{\mathrm{h}}|^3}\\-GM(|\bm{x}_{\mathrm{h}}|)\frac{\bm{x}_{\mathrm{h}}}
{|\bm{x}_{\mathrm{h}}|^3},
\end{multline}
where $\bm{x}_{\mathrm{h}}$ is the position of the host and $G$ is the gravitational constant. The first term is the
gravitational attraction by the dwarf galaxy, the second one is the attraction by the host galaxy and the last one is
the inertial force arising because the reference frame is not inertial, as the dwarf is falling onto the host.
The second term can be expanded in the series for $|\bm{x}_{\mathrm{s}}|/|\bm{x}_{\mathrm{h}}|\ll1$, yielding
\begin{multline}
\ddot{\bm{x}}_{\mathrm{s}}
=-Gm(|\bm{x}_{\mathrm{s}}|)\frac{\bm{x}_{\mathrm{s}}}{|\bm{x}_{\mathrm{s}}|^3}
-GM(|\bm{x_{\mathrm{h}}}|)\frac{\bm{x}_{\mathrm{s}}}{|\bm{x}_{\mathrm{h}}|^3}\\
+GM(|\bm{x}_{\mathrm{h}}|)[3-p(|\bm{x}_{\mathrm{h}}|)]\frac{(\bm{x}_{\mathrm{s}}
\cdot\bm{x}_{\mathrm{h}})\bm{x}_{\mathrm{h}}}{|\bm{x}_{\mathrm{h}}|^5}+\mathcal{O}
\left[(|\bm{x}_{\mathrm{s}}|/|\bm{x}_{\mathrm{h}}|)^2\right],
\end{multline}
where $p(|\bm{x}_{\mathrm{h}}|)=\mathrm{d}\log M/\mathrm{d}\log x|_{x=|\bm{x}_{\mathrm{h}}|}$ is the logarithmic
derivative of $M(x)$ calculated at $x=|\bm{x}_{\mathrm{h}}|$ and the dot denotes the scalar product.

The second and third terms together are commonly referred to as the tidal force. It squeezes the dwarf in the plane
perpendicular to the direction to the host and elongates it in the direction towards and away from the host. In
such an environment an initially disky dwarf is likely to form a tidally induced bar \citep[e.g.][]{lokas14}.

Let us change the reference frame to one rotating with a variable angular velocity $\bm{\Omega}$.
New terms, the inertial forces, appear: the Euler force, the Coriolis force and the centrifugal force.
The acceleration of the star now reads
\begin{multline}
\ddot{\bm{x}}_{\mathrm{s}}=-Gm(|\bm{x}_{\mathrm{s}}|)\frac{\bm{x}_{\mathrm{s}}}{|\bm{x}_{\mathrm{s}}|^3}
-GM(|\bm{x_{\mathrm{h}}}|)\frac{\bm{x}_{\mathrm{s}}}{|\bm{x}_{\mathrm{h}}|^3}\\
+GM(|\bm{x_{\mathrm{h}}}|)[3-p(|\bm{x}_{\mathrm{h}}|)]\frac{(\bm{x}_{\mathrm{s}}
\cdot\bm{x}_{\mathrm{h}})\bm{x}_{\mathrm{h}}}{|\bm{x}_{\mathrm{h}}|^5}
-\dot{\bm{\Omega}}\times\bm{x}_{\mathrm{s}}\\
-2\bm{\Omega}\times\dot{\bm{x}}_{\mathrm{s}}
-\bm{\Omega}\times(\bm{\Omega}\times\bm{x}_{\mathrm{s}})
+\mathcal{O}\left[(|\bm{x}_{\mathrm{s}}|/|\bm{x}_{\mathrm{h}}|)^2\right],
\end{multline}
where the cross denotes the vector product.

In order to proceed we need to make an assumption about the velocity of the star.
Let us assume the star follows a circular orbit around the dwarf with angular velocity $\bm{\Omega}_{\mathrm{s}}$,
so in the non-rotating reference frame centered on the dwarf its velocity is
\begin{equation}
\dot{\bm{x}}_{\mathrm{s}}=\bm{\Omega}_{\mathrm{s}}\times\bm{x}_{\mathrm{s}}.
\label{eq_velocity_nonrot}
\end{equation}
In the
rotating frame we have to subtract the velocity of the frame itself, therefore
\begin{equation}
\dot{\bm{x}}_{\mathrm{s}}=(\bm{\Omega}_{\mathrm{s}}-\bm{\Omega})\times\bm{x}_{\mathrm{s}}.
\label{eq_velocity_rot}
\end{equation}
This expression is substantially different from what was assumed by \citet{read06}
and we discuss this issue below. After substitution of the velocity we obtain
\begin{multline}\label{eq_acceleration}
\ddot{\bm{x}}_{\mathrm{s}}=-Gm(|\bm{x}_{\mathrm{s}}|)
\frac{\bm{x}_{\mathrm{s}}}{|\bm{x}_{\mathrm{s}}|^3}
-GM(|\bm{x}_{\mathrm{h}}|)\frac{\bm{x}_{\mathrm{s}}}
{|\bm{x}_{\mathrm{h}}|^3}\\+GM(|\bm{x_{\mathrm{h}}}|)
[3-p(|\bm{x}_{\mathrm{h}}|)]\frac{(\bm{x}_{\mathrm{s}}
\cdot\bm{x}_{\mathrm{h}})\bm{x}_{\mathrm{h}}}{|\bm{x}_{\mathrm{h}}|^5}-
\dot{\bm{\Omega}}\times\bm{x}_{\mathrm{s}}\\
-2\bm{\Omega}\times(\bm{\Omega}_{\mathrm{s}}
\times\bm{x}_{\mathrm{s}})+\bm{\Omega}
\times(\bm{\Omega}\times\bm{x}_{\mathrm{s}})
+\mathcal{O}\left[(|\bm{x}_{\mathrm{s}}|
/|\bm{x}_{\mathrm{h}}|)^2\right],
\end{multline}
which is the final formula for the acceleration of the star on a circular orbit around the dwarf galaxy,
calculated in the frame rotating with a variable angular velocity. We note that up to this point vectors
$\bm{\Omega}_{\mathrm{s}}$ and $\bm{\Omega}$ are arbitrary.

We now derive the tidal radius for stars whose orbits lie in the same plane as the orbit of the dwarf. We
choose $\bm{\Omega}$ so that the unit vector $\hat{\bm{x}}_{\mathrm{h}}=\bm{x}_{\mathrm{h}}/|\bm{x}_{\mathrm{h}}|$,
pointing from the dwarf towards the host, is constant in time. Such an $\bm{\Omega}$ is equal to the
instantaneous orbital angular velocity vector of the dwarf.
In such a setup
$\bm{\Omega}\parallel\bm{\Omega}_{\mathrm{s}}$ and $\bm{\Omega}\perp\bm{x}_{\mathrm{s}}$.
Using the identity $\bm{A}\times(\bm{B}\times \bm{C}) = (\bm{A}\cdot \bm{C})\bm{B}-(\bm{A}\cdot
\bm{B})\bm{C}$ and these properties we can rewrite double vector products obtaining
\begin{multline}
\ddot{\bm{x}}_{\mathrm{s}}=-Gm(|\bm{x}_{\mathrm{s}}|)\frac{\bm{x}_{\mathrm{s}}}
{|\bm{x}_{\mathrm{s}}|^3}-GM(|\bm{x_{\mathrm{h}}}|)\frac{\bm{x}_{\mathrm{s}}}
{|\bm{x}_{\mathrm{h}}|^3}\\+GM(|\bm{x_{\mathrm{h}}}|)[3-p(|\bm{x}_{\mathrm{h}}|)]
\frac{(\bm{x}_{\mathrm{s}}\cdot\bm{x}_{\mathrm{h}})\bm{x}_{\mathrm{h}}}
{|\bm{x}_{\mathrm{h}}|^5}-\dot{\bm{\Omega}}\times\bm{x}_{\mathrm{s}}\\
+2(\bm{\Omega}\cdot\bm{\Omega}_{\mathrm{s}})\bm{x}_{\mathrm{s}}-
|\bm{\Omega}|^2\bm{x}_{\mathrm{s}}+\mathcal{O}\left[(|\bm{x}_{\mathrm{s}}|
/|\bm{x}_{\mathrm{h}}|)^2\right].
\end{multline}

Stars are mainly stripped when they are at the smallest or largest distance from the host and leave the vicinity of the
dwarf through the L1 and L2 Lagrange points. To find the component of the acceleration along the direction towards the
host assuming $\bm{x}_{\mathrm{s}}\parallel\bm{x}_{\mathrm{h}}$ we multiply the above equation by
$\hat{\bm{x}}_{\mathrm{s}}=\bm{x}_{\mathrm{s}}/|\bm{x}_{\mathrm{s}}|$ to get
\begin{multline}\ddot{\bm{x}}_{\mathrm{s}}\cdot\hat{\bm{x}}_{\mathrm{s}}=
-Gm(x_{\mathrm{s}})\frac{1}{x_{\mathrm{s}}^2}+GM(x_{\mathrm{h}})
[2-p(x_{\mathrm{h}})]\frac{x_{\mathrm{s}}}{x_{\mathrm{h}}^3}\\
+\Omega^2\left(2 \frac{\Omega_{\mathrm{s}}}{\Omega}-1\right)
x_{\mathrm{s}}+\mathcal{O}\left[(x_{\mathrm{s}}/x_{\mathrm{h}})^2\right],
\end{multline}
where we denoted magnitudes of each vector as $|\bm{a}|=a$, except for $\Omega_{\mathrm{s}}$ which obeys the relation
$\bm{\Omega}\cdot\bm{\Omega}_{\mathrm{s}}=\Omega\Omega_{\mathrm{s}}$, i.e. its sign depends on its
direction with respect to the direction of $\bm{\Omega}$. For stars on prograde orbits with respect
to the orbital velocity of the dwarf $\Omega_\mathrm{s}>0$, while for stars on retrograde orbits
$\Omega_\mathrm{s}<0$.

We can parametrize the orbital angular velocity by the relation to the circular orbital angular velocity
of the host $\Omega_\mathrm{circ}(x_\mathrm{h})$ as
\begin{equation}
\Omega^2=\frac{\Omega_\mathrm{circ}^2(x_\mathrm{h})}{\lambda(x_{\mathrm{h}})}=\frac{GM(x_{\mathrm{h}})}{x_{\mathrm{h}}^3
\lambda(x_{\mathrm{h}})}.
\end{equation}
The parameter $\lambda(x_{\mathrm{h}})$ depends on the orbit of the dwarf and details of the
host potential. For a circular orbit $\lambda=1$, for a point-mass host
$\lambda(x_{\mathrm{h}})=x_{\mathrm{h}}/[a(1-e^2)]$, where $a$ is the major semi-axis and $e$ is the eccentricity
of the orbit. One may wonder if we can further assume something about $\Omega_{\mathrm{s}}$, for example that it
is equal to the circular
velocity in the dwarf, i.e. $\Omega_{\mathrm{s, circ}}=\pm\left[Gm(x_{\mathrm{s}})/x_{\mathrm{s}}^3\right]^{1/2}$.
However, such an assumption leads to incorrect results, as we demonstrate below.

The tidal radius $r_{\mathrm{t}}$ is commonly defined as the distance from the center of the dwarf where there is
neither acceleration towards the dwarf nor towards the host \citep[e.g.][]{king62, read06}. In our case this can be
written as $\ddot{\bm{x}}_{\mathrm{s}}\cdot\hat{\bm{x}}_{\mathrm{s}}|_{x_{\mathrm{s}}=r_{\mathrm{t}}}=0$. After
applying this relation, substituting for $\Omega^2$ and truncating the series we obtain the equation for $r_{\mathrm{t}}$
\begin{gather}
-Gm(r_{\mathrm{t}})\frac{1}{r_{\mathrm{t}}^2}+GM(x_{\mathrm{h}})
[2-p(x_{\mathrm{h}})]\frac{r_{\mathrm{t}}}{x_{\mathrm{h}}^3}+ \\
+\frac{GM(x_{\mathrm{h}})}{x_{\mathrm{h}}^3
\lambda(x_{\mathrm{h}})}\left(2 \frac{\Omega_{\mathrm{s}}}{\Omega}-1\right)
r_{\mathrm{t}}=0
\end{gather}
which can be rewritten in a more convenient form
\begin{gather}\label{eq_tidal_radius}
r_{\mathrm{t}}=
x_{\mathrm{h}}\left(\frac{[m(r_{\mathrm{t}})/
M(x_{\mathrm{h}})]\lambda(x_{\mathrm{h}})}{2\Omega_{\mathrm{s}}/
\Omega-1+[2-p(x_{\mathrm{h}})]\lambda(x_{\mathrm{h}})}\right)^{1/3},\\
2\Omega_{\mathrm{s}}/\Omega>-[2-p(x_{\mathrm{h}})]\lambda(x_{\mathrm{h}})+1.
\end{gather}
The above inequality has to be fulfilled, otherwise the tidal radius does not exist. In such a situation the acceleration
points towards the dwarf. One can interpret this as the tidal radius being infinite. However, we assumed that
$r_{\mathrm{t}}\ll x_{\mathrm{h}}$, hence such an extrapolation is not valid.

If we assume the simplest setup, namely a point-mass host (so $M(x_{\mathrm{h}})=M$ and $p=0$)
and dwarf ($m(r_{\mathrm{t}})=m$), and a circular orbit of the dwarf
($\lambda=1$), we obtain
\begin{gather}
r_{\mathrm{t}}=x_{\mathrm{h}}\left(\frac{1}{2\Omega_{\mathrm{s}}/
\Omega+1}\frac{m}{M}\right)^{1/3},
\label{eq_simple_rt}
\\ \Omega_{\mathrm{s}}/\Omega>-1/2.
\end{gather}
Assuming further that $\Omega_{\mathrm{s}}/\Omega=1$ we recover the expression for the L1/L2 Lagrange
points, sometimes also referred to as the Jacobi radius $r_{\mathrm{J}}$ \citep[see e.g.][]{bt08}:
\begin{equation}
r_{\mathrm{t}}=r_{\mathrm{J}}=x_{\mathrm{h}}\left(\frac{m}{3M}\right)^{1/3}.
\label{eq_jacobi}
\end{equation}
This is straightforward to understand: a star located at the L1/L2 point is stationary in the rotating reference frame,
hence it rotates around the dwarf with the same angular velocity as the dwarf revolves around the host.

From the above it is also clear why we cannot assume the star to be moving with angular velocity
$\Omega_{\mathrm{s}}=[Gm(x_{\mathrm{s}})/x_{\mathrm{s}}^3]^{1/2}$. In such case the angular velocity at the true
distance of the L1/L2 points would be $\Omega_{\mathrm{s}}=\sqrt{3}\Omega$, thus significantly overestimated.
Consequently, the tidal radius would be underestimated. Therefore, as $\Omega_{\mathrm{s}}$ one has to use the
angular velocity \emph{already perturbed} by the host.

It may seem that for a dwarf on a retrograde orbit ($\Omega_{\mathrm{s}}<0$), which spins sufficiently fast, the
tidal radius is infinite. However, as we have already argued, in this case our approximation breaks down. Moreover, in
real objects $|\Omega_\mathrm{s}|$ decreases with the distance form the center, therefore there will exist a radius
where the necessary inequality will be fulfilled.

If we assume in equation \eqref{eq_tidal_radius} that the star is at the (instantaneous) Lagrange point
($\Omega_{\mathrm{s}}/\Omega=1$) we can recover two known expressions for the tidal radius. For the dwarf on a
circular orbit ($\lambda=1$) around an extended host we recover the formula 8.108
from \citet{bt08}, namely
\begin{equation}
r_{\mathrm{t}}=x_{\mathrm{h}}\left[\frac{1}{3-p(x_{\mathrm{h}})}\frac{m(r_{\mathrm{t}})}{M(x_{\mathrm{h}})}\right]^{1/3}.
\end{equation}
For an elliptical orbit ($\lambda=x_{\mathrm{h}}/[a(1-e^2)]$) around a
point-mass host ($p=0$) we recover the limiting radius formula (7) from \citet{king62}:
\begin{equation}
r_{\mathrm{t}}=
x_{\mathrm{h}}\left[\frac{m(r_{\mathrm{t}})/M}{ (a/x_{\mathrm{h}})(1-e^2) +2}\right]^{1/3}.
\end{equation}
We emphasize that the
exceptionally simple formula $r_{\mathrm{t}}=x_{\mathrm{h, peri}}\left\{m/[M(3+e)]\right\}^{1/3}$ is valid only
at the pericenter
of the satellite's orbit (i.e. for $x_{\mathrm{h}} = x_{\mathrm{h, peri}} = a(1-e)$).

Let us now comment on the differences between our derivation and the one presented in \citet{read06}.
Our formulae for the star velocity \eqref{eq_velocity_nonrot} and \eqref{eq_velocity_rot}
should correspond to what \citet{read06}
assumed in their equation (3).
In the non-rotating reference frame centered on the dwarf their expression is equivalent to
\begin{equation}
\dot{\bm{x}}_{\mathrm{s}}=(\bm{\Omega}_{\mathrm{s}}+\bm{\Omega})\times\bm{x}_{\mathrm{s}}.
\end{equation}
This should be directly compared to our formula \eqref{eq_velocity_nonrot}.
\citet{read06} also assume that $\Omega_\mathrm{s}=\alpha\Omega_\mathrm{s,circ}$,
where $\Omega_\mathrm{s,circ}$ is the circular angular velocity of the dwarf
and $\alpha=1$, $0$ or $-1$ accounts for, respectively, prograde, radial and retrograde orbits of the stars.
Hence, according to \citet{read06}, when the star is located at the tidal radius,
it has an effective angular velocity of $\alpha\Omega_{\mathrm{s,circ}}+\Omega$.
It means that the stars on prograde and radial orbits are accelerated by a constant
value, equal to the orbital angular velocity of the dwarf.
The stars on retrograde orbits are first decelerated and then accelerated, but
then they orbit in a different direction than the body of the dwarf.
The radial orbits, initially having $\Omega_\mathrm{s}=0$, are in \citet{read06}
framework accelerated to $\Omega_\mathrm{s}=\Omega$.
Thus it is not surprising that the formula of \citet{read06} reduced for the radial orbits to
the one of \citet{king62}, as equality of the angular velocities is precisely the assumption of the latter.

From the tidal stirring simulations of dwarfs on prograde orbits it is known that the rotation of
the dwarf diminishes, rather than increases \citep[see e.g.][]{lokas15}. This explains why \citet{lokas13}
obtained good match of the
tidal radius to $N$-body simulations assuming radial orbits of the stars, whereas the dwarfs were in
fact on mildly prograde orbits. Simply, the stars did not rotate as fast as \citet{read06} assumed
in the prograde case. We discuss this issue further in the next sections.

\section{Comparison to $N$-body simulations}
\label{sec_comparison}

Any formula should be compared to a realistic situation in order to check its validity and the assumptions
made during the derivation. We shall use two of the $N$-body simulations studied in detail by \citet{lokas15} that
followed the evolution of a dwarf galaxy around a Milky Way-like host.
The live realization of a Milky
Way-like galaxy consisted of an NFW \citep{nfw95} dark matter halo with a virial mass of $7.7\times10^{11}$~M$_\sun$ and
concentration $c=27$ and an exponential disk of mass $3.4\times10^{10}$ M$_{\sun}$ with a radial scale length $2.88$ kpc
and thickness $0.44$ kpc. The dwarf galaxy had an NFW halo of $10^9$ M$_{\sun}$ and $c=20$ and a disk
of mass $2\times10^7\mathrm{M}_{\sun}$, the radial scale length $0.41$ kpc and thickness $0.082$ kpc. Each
component of each galaxy contained $10^6$ particles.

The dwarf galaxy was initially placed at an apocenter of an eccentric orbit around the host galaxy with apocenter and
pericenter distances of $120$ and $25$ kpc, respectively. The plane of the orbit was the same as the plane of the
host's disk. The only parameter which was varied between the simulations was the initial inclination of the
disk of the dwarf with respect to the orbit. Here we will use only two configurations: the exactly prograde and
the exactly retrograde, which in \citet{lokas15} were referred to as I0 and I180, respectively.

In principle, the tidal radius formula \eqref{eq_tidal_radius} applies to individual particles in the dwarf, as each
star has a different angular velocity $\Omega_\mathrm{s}$. In order to apply it to the dwarf as a whole, we
measured the $\Omega_\mathrm{s}(x_\mathrm{s})$ profile, i.e. the profile of the mean angular velocity component
parallel to the orbital angular velocity of the dwarf. Obviously, particles in the dwarf are not on exactly circular
orbits and have non-negligible radial velocities. As a result, an additional component of the Coriolis force appears.
However, it is orthogonal to $\bm{x}_\mathrm{s}$ and, consequently, does not have any impact on the force component
along $\bm{x}_\mathrm{s}$.

\begin{figure}
\includegraphics[width=\columnwidth]{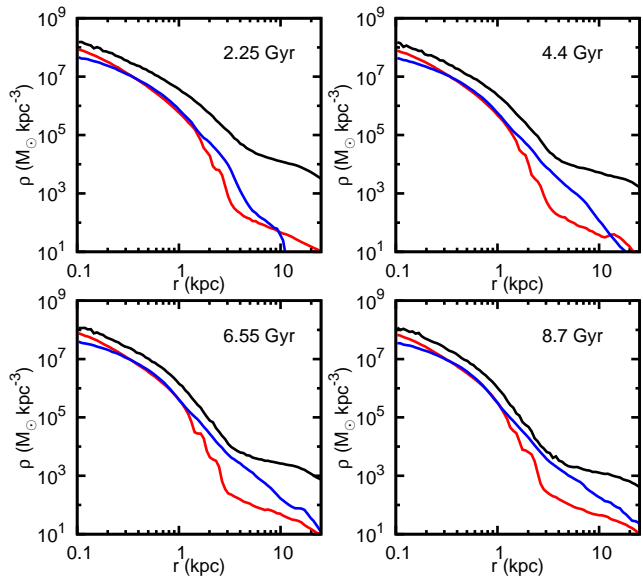}
\caption{Density profiles of the stellar component of the dwarf galaxy for the prograde model (red) and the retrograde
model (blue), as well as dark matter density profile (black) at consecutive apocenters.}
\label{fig_density_profiles}
\end{figure}

The right-hand side of equation \eqref{eq_tidal_radius} depends on $r_\mathrm{t}$, thus it is an implicit
formula for the tidal radius, which can be solved numerically. In order to obey the assumptions, from
now on we will treat the disk of the host as a point mass. Furthermore, we calculated $m(r_\mathrm{t})$ assuming
spherical symmetry of the dwarf galaxy mass distribution. While these approximations introduce some error, we
consider it negligible, as it only underestimates the gravitational force of the disks of both galaxies, which are at
least an order of magnitude less massive than their dark matter haloes. The calculation of
$\lambda(x_\mathrm{h})$ was carried out assuming that the dwarf is a point mass object orbiting a spherical host.

\begin{figure}
\includegraphics[width=\columnwidth]{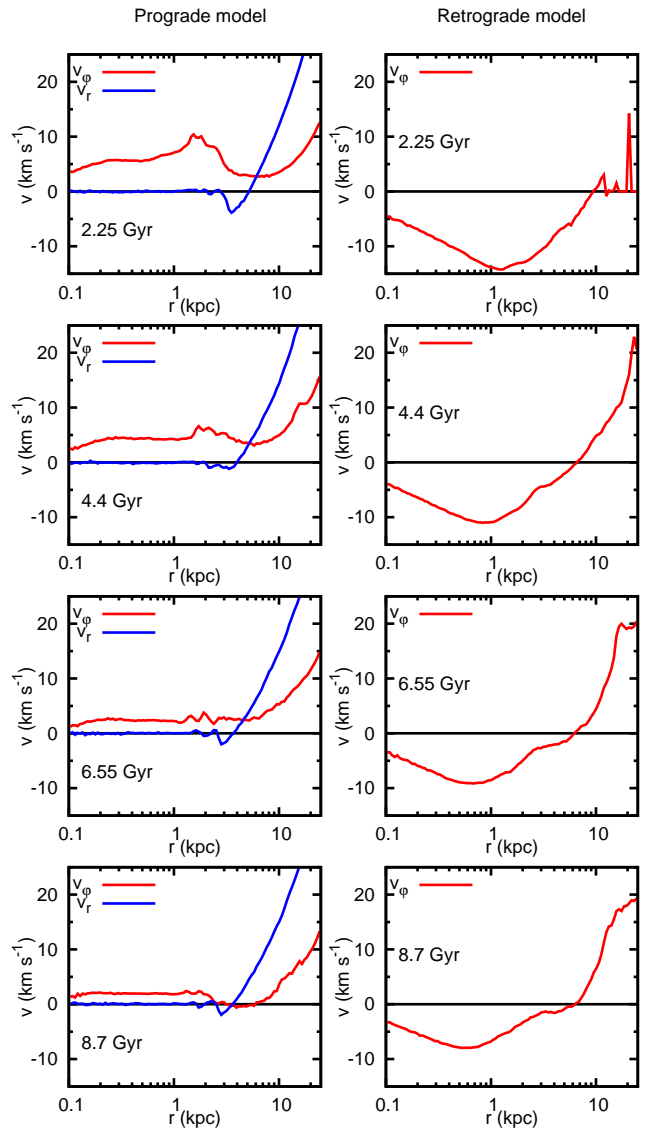}
\caption{\emph{Left column:} The rotational velocity (red) and the radial velocity (blue) of the dwarf on the prograde
orbit. \emph{Right column:} The rotational velocity of the dwarf on the retrograde orbit. Negative values indicate that
the dwarf counter rotates with respect to its orbit. Rows of panels correspond to consecutive apocenters.}
\label{fig_velocity}
\end{figure}

In Figure \ref{fig_density_profiles} we show density profiles of the stellar component during consecutive
apocenters for both models. In addition, we plot the profile of the dark matter component, which was very similar
in both
cases. In the case of the prograde model there is a clear break in the density profiles
where the main body of the dwarf ends and
the tidal tails begin. On the other hand, the profiles of the retrograde model are rather featureless and close to a
power law.

For their mildly prograde models, \citet{lokas13} compared the break radius $r_\mathrm{b}$ to the theoretical tidal
radius. Here, we were only able to locate the break in the prograde model and to measure it we used the method of
\citet{johnston02}. At each radius we calculated slopes of the density profile interior and exterior to that radius.
Comparing the two slopes we searched for the innermost difference larger than the assumed threshold.

Stripping of the stars should also influence their kinematics. This is because in the first approximation
the stars bound to the dwarf are mostly influenced by the dwarf potential, whereas the motion of the stripped ones is
governed by the potential of the host. Therefore, we introduce the \emph{kinematic radius} $r_\mathrm{k}$ beyond which
the kinematics of the particles is no longer dominated by the potential of the dwarf.

Particular realizations of this definition are different in both models. To illustrate this, in Figure
\ref{fig_velocity} we plot profiles of the rotational velocity $v_{\varphi}=r\Omega_\mathrm{s}$ of the dwarfs at the
consecutive apocenters. The inner parts of the retrograde model counter-rotate with respect to the orbital angular
velocity, as indicated by the negative values of $v_{\varphi}$. The maximum of the velocity curve slowly decreases
at subsequent apocenters due to mass loss. Particles which were stripped and are already
far away (more than $10$ kpc) have positive angular velocity with respect to the center of the dwarf, as they orbit the
host. Taking this into account, we define the kinematic radius as the one where $\Omega_\mathrm{s}=0$.

In the case of the prograde model, the angular velocity in the inner parts decreases significantly as the dwarf is
transformed and the random motions start to dominate \citep[see e.g. Fig. 2 in][]{lokas15}. In the
outer parts, $v_{\varphi}$ grows similarly as in the retrograde model. Unfortunately, there is no such a striking
feature as in the retrograde case. One can see a minimum in the outer parts, but it is very wide and shallow,
if present at all, therefore it is not a suitable proxy for the kinematic radius. Instead, we define
$r_\mathrm{k}$ as a radius where the radial velocity is equal to the rotational velocity, i.e. $v_r=v_{\varphi}$.
We note that there is a region in the profiles of $v_r$ where $v_r<0$, indicating that some particles are
reaccreted onto the dwarf.

\begin{figure}
\includegraphics[width=\columnwidth]{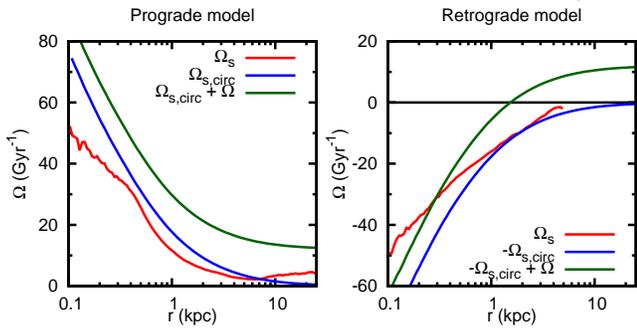}
\caption{The angular velocity profile of stars (red), the circular angular
velocity curve (blue) and the effective angular velocity (green), as assumed by \citet{read06}.
The left panel is for the prograde and the right one for the retrograde model. Negative values
in the retrograde model indicate counter-rotation with respect to the orbital angular velocity.
Both panels show measurements from the simulations at the first pericenter passage, after $1.15$ Gyr of evolution.}
\label{fig_read06}
\end{figure}

Let us now assess the assumption of \citet{read06} concerning the effective angular velocity of the stars in the dwarf.
In Figure \ref{fig_read06} we compare the profiles of the measured
angular velocity of the dwarf $\Omega_s$, the circular angular velocity $\Omega_\mathrm{s,circ}$ and the
effective angular velocity $\alpha\Omega_\mathrm{s,circ}+\Omega$ assumed by \citet{read06}. The measurements were done
at the first pericenter passage.
Near the centers,  the $\Omega_\mathrm{s}$ profile is shallower than $\Omega_\mathrm{s,circ}$ partially
due to the former being measured in spherical bins.
It is well visible that the assumption of \citet{read06} is not valid, as it significantly overestimates
the angular velocity, which increases the Coriolis force term and leads to a large underestimation of the tidal radius.
The magnitude of $\Omega_\mathrm{s}$ is smaller than $\Omega_\mathrm{s,circ}$ even at the first pericenter,
and later on the discrepancy is even larger. Thus, as we discussed previously, $\Omega_\mathrm{s,circ}$
cannot be used as a proxy for $\Omega_\mathrm{s}$.

\begin{figure}
\includegraphics[width=\columnwidth]{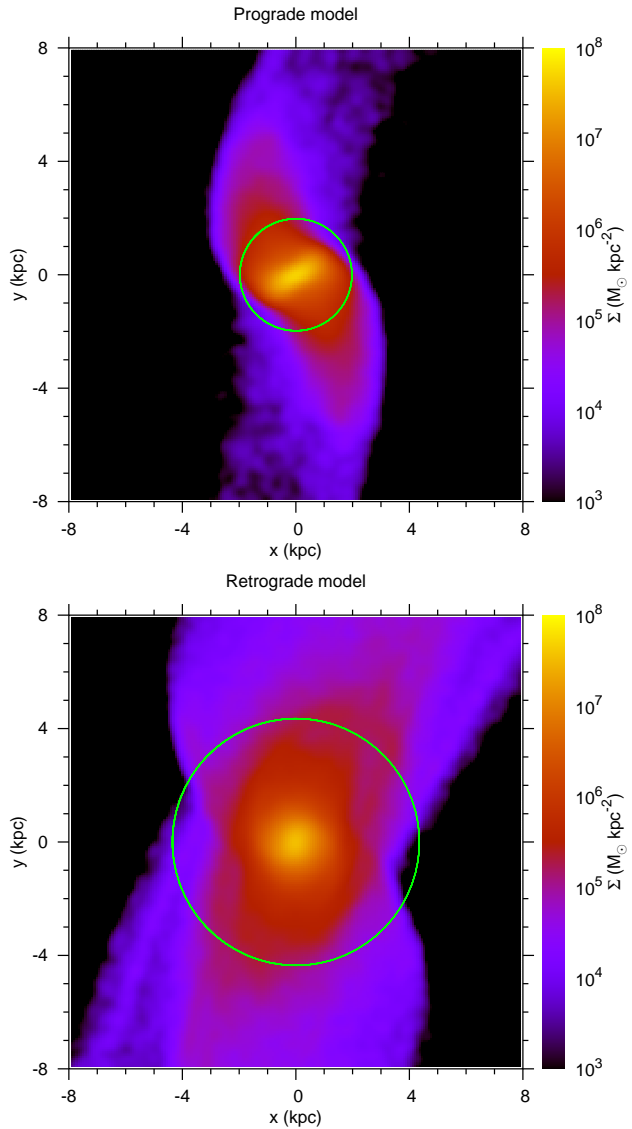}
\caption{
Surface density maps of the stellar component of dwarf galaxies just after the third pericenter passage ($5.5$ Gyr).
The green circles indicate the tidal radii calculated from formula \eqref{eq_tidal_radius}.
The top panel is for the prograde and the bottom one for the retrograde model.
}
\label{fig_map}
\end{figure}

Figure \ref{fig_map} illustrates the differences between the distributions of the stellar component
in both models and their relation to the tidal radius calculated from our equation \eqref{eq_tidal_radius}.
During the third pericenter passage, the dwarf on the prograde orbit is already significantly smaller
than the one on the retrograde orbit. In the prograde orbit case, the material is also more effectively
removed from the vicinity of the dwarf. The calculated tidal radii correspond to the actual sizes of the dwarfs.

\begin{figure}
\includegraphics[width=\columnwidth]{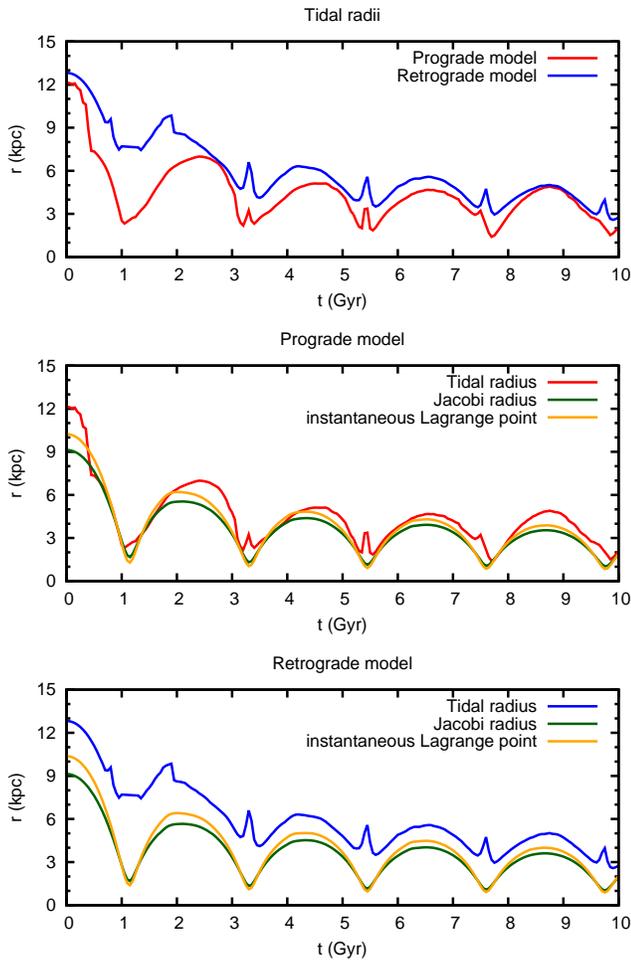}
\caption{\emph{Top:} Tidal radii calculated for the prograde (red) and the retrograde (blue) model. \emph{Middle:}
Tidal radius (red), Jacobi radius (green) and instantaneous Lagrange point (orange) for the prograde model. \emph{Bottom:}
Tidal radius (blue), Jacobi radius (green) and instantaneous Lagrange point (orange) for the prograde model.}
\label{fig_radius_teo}
\end{figure}

In Figure \ref{fig_radius_teo} we show tidal radii calculated from formula \eqref{eq_tidal_radius} and
compare them to tidal radii calculated with different assumptions. Let us recall that the Jacobi formula
\eqref{eq_jacobi} is the most widely used expression for the tidal radius and corresponds to the
distance of L1/L2 Lagrange points, as if the dwarf was on a circular orbit of radius equal to
its current distance from the host. In the instantaneous Lagrange point approximation we
assume $\Omega_\mathrm{s}/\Omega=1$ and use our formula \eqref{eq_tidal_radius}.
This is equivalent to the \citet{king62} approximation, only without the assumption of a point-mass host.

The tidal radius for the retrograde model is always larger than for the prograde model. However, the difference is
smaller later on because the rotation of the prograde dwarf slows down with time. Small peaks in $r_\mathrm{t}$
measurements around the pericenters are caused by the Coriolis force originating from the movement of the whole dwarf on
such an elongated orbit that locally exceeds the tidal force. This behavior might be different if we included higher order terms in the tidal force.

In the case of the prograde model, the simpler approximations give similar results. This is due to the fact
that the ratio $\Omega_\mathrm{s}/\Omega$ varies in a range from about $0.3$ to $2$, so the approximation of
the instantaneous Lagrange point is not that far from reality. The Jacobi radius performs worse, as it assumes
an instantaneous circular orbit of the satellite.
However, in the case of the retrograde model the difference is huge and results from taking into
account the orientation of the dwarf's rotation.

\begin{figure}
\includegraphics[width=\columnwidth]{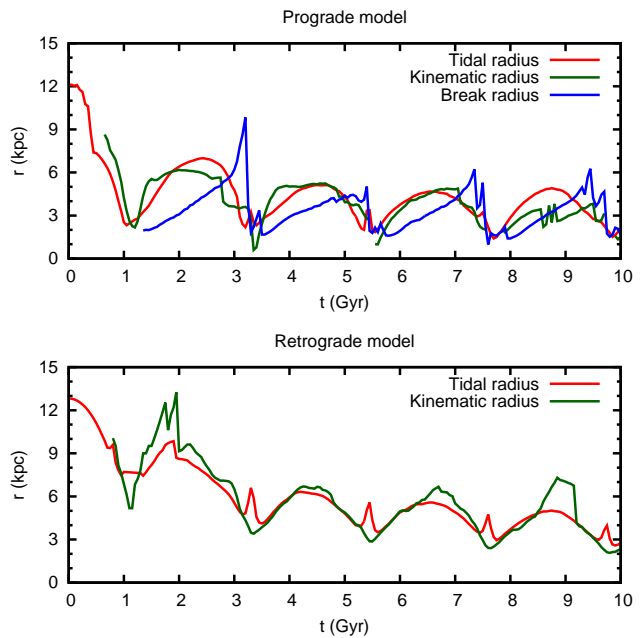}
\caption{\emph{Top:}
Tidal radius (red), kinematic radius (green) and break radius (blue) for the prograde model. \emph{Bottom:}
Tidal radius (red) and kinematic radius (green) for the retrograde model.}
\label{fig_radius_comp}
\end{figure}

In Figure \ref{fig_radius_comp} we compare our tidal radius with the kinematic and break radii
obtained from $N$-body simulations for the prograde and retrograde model.
The kinematic radius agrees remarkably well with the tidal radius. However, the agreement is worse between the fourth
and fifth pericenter. In the prograde case it is because the stellar component has already significantly evolved from
the initial disky configuration. In the retrograde case, the profile of the angular velocity in the outer parts is very
flat and it does not cross zero up to a large radius, possibly due to numerical errors. On the other hand,
$r_\mathrm{b}$ is only a rough estimate of the tidal radius. As discussed by \citet{lokas13}, it is because the density
distribution needs time to respond to the external forces.

\section{Conclusions}
\label{sec_conclusions}

In this work we studied the tidal radius of satellites orbiting in a host potential. We derived an improved formula for
$r_\mathrm{t}$ for stars revolving around the satellite in the same plane as the satellite orbits the host. Our
formula \eqref{eq_tidal_radius} properly takes into account that the star might be on a prograde or retrograde orbit.
We verified our formula against $N$-body simulations of a disky dwarf galaxy on a co-planar orbit around a Milky
Way-like host. The tidal radius corresponds to the kinematic radius, i.e. the radius beyond which kinematics of the
stellar particles is dominated by the host potential and the particles no longer orbit the dwarf galaxy.

We showed that the assumption made by \citet{read06} concerning the effective angular velocity of the stars in the dwarf
is not fulfilled in the case of an elongated orbit of the dwarf. Unfortunately, no justification of the used assumption is presented in the discussed paper.
\citet{read06} claim that for short timescales they
obtained excellent agreement between their formula and simulations. However, at their tidal radii one can notice
only some depletion of stars, but certainly not a boundary of the system. Thus, in our opinion, their tidal radius
is significantly underestimated. Moreover, they take into account particles on polar orbits, which are not well
described by either their or our formalism. 

As already noticed by \citet{lokas13}, the pericentric tidal radius solely, as calculated from e.g. \citet{king62} simple formula, is not enough to describe the extent of the satellite as it orbits around the host object. As both measured quantities, the break and the kinematic radius, show, the size of the satellite changes along the orbit. After the pericenter passage, the dwarf expands again (see Figure \ref{fig_radius_comp}). Therefore, the \citet{king62} conjecture that the satellites are trimmed at the pericenter and then remain unchanged, is not valid. Hence, to describe properly the size of a satellite one should use more advanced prescription.

Applying our formula \eqref{eq_tidal_radius} to the observations may be difficult, as one needs to know the
orbit of the satellite. However, this is also needed if one wants to use \citet{king62} expression.
As an improvement with respect to the classical formula for the Jacobi radius \eqref{eq_jacobi} one can use formula
\eqref{eq_simple_rt}, which holds under similar assumptions as the former (i.e. a circular orbit and a point-mass host),
but takes into account the rotation of the satellite, which can be readily measured for many objects. In this simple approximation, as the orbital angular velocity one may assume the circular orbital angular velocity.

Our derivation depends on two basic assumptions: the stars orbit the dwarf on circular orbits and they are stripped
in the vicinity of the line connecting the dwarf and the host (i.e. near L1/L2 Lagrange points). While the
latter seems to be fulfilled based on the analysis of tidal tails \citep{klimentowski09,lokas13}, the former is quite
disputable. For example, \citet{szebehely67} showed that in the restricted three-body problem prograde orbits of the
smallest body get complicated when particles approach the Lagrange points. On the other hand, retrograde orbits
are fairly circular. Thus, we expect that our calculations are more robust in the retrograde case.

Using an improved impulse approximation, \citet{donghia09, donghia10} proposed that stripping of the satellites is of
resonant origin. Namely, for prograde encounters particles have largest velocity gains if
their $\alpha_\mathrm{res}=\Omega_\mathrm{s}/\Omega \sim 1$ during the pericenter passage.
\citet{lokas15} calculated radii $r_\mathrm{res}$ at which $\alpha_\mathrm{res}=1$ for the
prograde model. In the vicinity of the pericenter, where most of the stripping occurs, $r_\mathrm{res}<r_\mathrm{t}$,
i.e.\ resonant radius is smaller than the tidal radius.

In view of the above, we may interpret the mechanism of stripping in the case of coplanar encounters as follows. In
the prograde case the material in the outer parts of the disk is accelerated and pushed away by the resonant mechanism.
When it travels beyond the tidal radius it becomes unbound from the dwarf and starts to follow the potential of the
host. Some of this material is reaccreted by the dwarf, as its tidal radius increases when the dwarf recedes from the
pericenter. In the retrograde case stellar particles are accelerated much more weakly (see Fig.\@ 4 of
\citealt{donghia10}), so they are driven out of the dwarf, but they remain in its vicinity, as confirmed by their
density profiles (Fig.\@ \ref{fig_density_profiles}).

Our analytic calculations are based on simplifying assumptions, which are not able to grasp other possible inclinations
of the dwarf's disk and the variety of particle orbits. More investigation is needed to establish a proper connection
between the stability of individual particle orbits and the general behavior of the density distribution and stripping.

\acknowledgments
This work was supported by the Polish National Science Centre under
grant 2013/10/A/ST9/00023. We acknowledge discussions with E. D'Onghia.

\end{document}